Development of spatial suppression surrounding the focus of visual attention

Audrey M. B. Wong-Kee-You[1], John K. Tsotsos[2,3], & Scott A. Adler[1,2]

[1]Department of Psychology, York University, Toronto, Canada; [2]Centre for Vision Research, Toronto, Canada; [3]Department of Electrical Engineering and Computer Science, Toronto Canada.

**Corresponding Author:** Audrey M. B. Wong-Kee-You, Department of Psychology, York University, Toronto, Canada. Email: audrey.wky@gmail.com

**Abstract**

The capacity to filter out irrelevant information from our environment is critical to efficient processing. Yet, during development, when building a knowledge base of the world is occurring, the ability to selectively allocate attentional resources is limited (e.g., Amso & Scerif, 2015). In adulthood, research has demonstrated that surrounding the spatial location of attentional focus is a suppressive field, resulting from top-down attention promoting the processing of relevant stimuli and inhibiting surrounding distractors (e.g., Hopf et al., 2006). It is not fully known, however, whether this phenomenon manifests in development. Could limitations in attentional focus in development be accounted for by reduced attention surround suppression, or ineffective top-down attentional modulation? In the current study, we examined whether spatial suppression surrounding the focus of visual attention is exhibited in developmental age groups. Participants between 12 and 27 years of age exhibited spatial suppression surrounding their focus of visual attention. Their accuracy increased as a function of the separation distance between a spatially cued (and attended) target and a second target, suggesting that a ring of suppression surrounded the attended target. When a central cue was instead presented and therefore attention was no longer spatially cued, surround suppression was not observed, indicating that our initial findings of suppression were indeed related to the focus of attention. Attentional surround suppression was not observed in 8- to 11-years-olds, even with a longer spatial cue presentation time, demonstrating that the lack of the effect at these ages is not due to slowed attentional feedback processes. Our findings demonstrate that top-down attentional processes are still immature until approximately 12 years of age, and that they continue to be refined throughout adolescence, converging well with previous research on attentional development. Our findings, however, uniquely demonstrate that attentional surround suppression, a predicted by-product of top-down modulation of visual processing, is observed in in pre-adolescence and adolescence but not in childhood.

**Significance**

Attention is undoubtedly vital because without our brain's ability to filter relevant information from the overabundance of all available information, we would not be able to interpret and make sense of our environment. In development, attention is even more critical because it is a time period during which an immense amount of learning and psychological change is taking place. Understanding the functioning of visual attention processes in younger age groups and how these processes change over development is therefore critical. The current study is an important step demonstrating that top-down attention similarly affects visual processing from pre-adolescence to young adulthood, while additionally highlighting how visual attention processes function differently in childhood.

In our environment, there is an overabundance of available visual information. Our visual system has a limited processing capacity, and as a result it cannot process all the information it receives from our eyes (Carasco, 2011). Our brains must instead use attention to bring important information into focus, while filtering out irrelevant information (Driver, 2001). Attention mechanisms are understood to involve the interaction of specific neural systems that allow for the control of information processing and action (Hopf et al., 2012). Within the visual domain, attention mechanisms operate on different visual representations, such as spatial or location-, feature-, and object-based representations (Hopf et al., 2012). Regardless of the visual representations upon which it is operating, however, the functional consequence of attention mechanisms are believed to be the optimization of the visual system (Tsotsos, 2011; Carrasco, 2011).

But how does attention optimize the visual system or optimize the processing of visual information? Within the spatial domain, previous animal studies have revealed direct evidence that the focus of spatial attention impacts activity in early and intermediate visual areas of the brain, thereby facilitating the processing of relevant visual information (Sunberg et al., 2009; Zhang et al., 2014). But, perplexingly, greater levels of suppression are also found for stimuli immediately surrounding the focus of attention than for stimuli that are further away (Sundberg et al., 2009). This phenomenon of suppression surrounding the focus of attention is in fact a prediction of the selective tuning (ST) model of attention (Tsotsos, 1995).

According to the ST model, top-down attentional selection prunes and suppresses forward-projecting units or neurons not representing relevant input, which leads to enhanced processing of the attended input, but as a consequence also gives rise to spatial suppression surrounding the focus of attention (Tsotsos, 2005). The ST model views the visual processing architecture as a hierarchical and layered pyramid in which units or neurons within the network receive both feedforward (bottom-up) and feedback (top-down) connections. A winner-take-all[1] (WTA) process initially localizes the neurons with the largest response at the top layer. All of the connections of the neurons that do not contribute to the winner are inhibited. This strategy of finding the winners, layer by layer, and then pruning away irrelevant connections is applied recursively. The remaining connections can be considered as the pass zone or the spotlight of attentional focus, while the pruned connections form the suppressive surround. Neurally, the sources of top-down attentional signals are hypothesized to be a network of frontoparietal regions (Zanto & Rissman, 2015), including the frontal eye fields (Couperus & Mangun, 2010; Seiss, Driver & Eimer, 2009), inferior frontal junction (IFJ) (Sylvester, Jack, Corbetta & Shulman, 2008), superior frontal and angular gyri (Ruff & Driver, 2006), and precuneus (Payne & Allen, 2011).

Several studies have provided psychophysical (e.g., Cutzu & Tsotsos, 2003) and neural evidence of surrounding spatial suppression in adult humans (e.g., Boehler et al., 2009). For instance, Cutzu and Tsotsos (2003) had participants discriminate between two target letters. Spatial attention was cued to one of the targets. Participants' accuracy increased as a function of inter-target separation, suggesting that a surround suppressive ring accompanied the processing advantage allocated by the spatial cue. In a magnetoencephalography (MEG) study by Hopf and colleagues (2006), it was found that the MEG response was significantly reduced when a target appeared at a position next to where attention was allocated, suggesting that in the immediate surround of the focus of attention, is a region of suppression or neural attenuation. Though these studies demonstrate that attentional surround suppression is observed in human adults, it is not fully known whether this phenomenon is exhibited in development. The goal of the current study, therefore, was to examine whether spatial suppression surrounding the focus of visual attention is exhibited in younger age groups and, if so, to determine its development course. By examining when in development attentional suppression is observed, we also intended to examine the effectiveness of top-down attentional modulation across development.

[1] Winner-take-all is a parallel algorithm that localizes the maximum value of a set. (Koch & Ullman, 1985)

Notably, studies focusing on the development of top-down (feedback, intentional or goal-driven) or bottom-up (feedforward, reflexive) attentional processes have revealed differences in the maturation timeline of these processes. Visual search studies, for instance, have shown that bottom-up attentional processes are mature quite early in development, but that top-down processes are still developing in childhood. In difficult cases where the target shares features with the distractors, as in a conjunction search, children up to about 6 to 7 years of age have difficulty searching for the target (Donnelly, Cave, Greenway, Hadwin, Stevenson, & Sonuga-Barke, 2007; Trick & Enns, 1998; Woods, Göksun, Chatterjee, Zelonis, Mehta & Smith, 2013). Under conditions where the target is more salient, however, and obviously different from distractors, young children can accurately search and locate a target much like adults (Donelly et al., 2007; Merrill & Conners, 2013; Taylor, Chevalier& Lobaugh 2003; Trick & Enns, 1998; Woods et al., 2013). Studies using different tasks have also revealed findings that confirm the interpretation of late developing top-down attentional processes. For example, children have been found to be more vulnerable to capture by irrelevant stimuli than adults, presumably because their top-down attentional processes are still developing (Gaspelin et al., 2015).

Models and frameworks of attentional development propose that early in development visual feedfoward and low-level orienting processes are more dominant and as development progresses top-down feedback processes are strengthened (Amso & Scerif, 2005; Johnson, 1990;

Aktinson, 2002). Studies on brain development have also pointed to differences in the maturation timeline of low-level feedforward (bottom-up) and feedback (top-down) processes. For instance, the organization of cortical long-range connections involving increasingly frontal cortical areas continues to develop into childhood and adolescence (Fair, Cohen, Power, Dosenbach, Church, Miezin, Schlaggar & Petersen, 2009; Sepekar, Musen & Menon, 2009). Increases in myelination and white matter integrity that facilitates long-range communication, continue late in development (Raznahan, Shaw, Lerch, Clasen, Greenstein, Berman, Pipitone, Chakravarty & Giedd, 2014; Vandekar, Shinohara, Raznahan, Roalf, Ross, DeLeo, Ruparel, Verma, Wolf, Gur, & Gur, 2015).

Imaging studies specifically examining the development of attention networks also support the notion of late developing top-down processes. There are two partially segregated attention networks in the human brain: the dorsal and ventral attention networks (Corbetta & Shulman, 2002). Each network includes different brain areas that are believed to play a different role in attention. The dorsal attention network (DAN) shows activation when attention is focused, and is believed to be responsible for goal-driven top-down processing (Corbetta & Shulman, 2002). The ventral attention network (VAN) is generally activated in cases where bottom-up processing is active, such as when an unexpected event occurs and breaks an observer's attention from a given task (Corbetta & Shulman, 2002). Of relevance, the frontoparietal regions in the DAN are believed to be sources of attention biases onto the sensory cortex (i.e., visual cortex) (Desimone & Duncan, 1995; Corbetta & Shulman, 2002; Reynolds & Chelazzi, 2004), and therefore likely play an important role in the presentation of suppression surrounding the focus of attention.

In a recent study, Farrant and Uddin (2015) used resting state fMRI to examine the development of DAN and VAN in children aged between 7 and 12 years. Farrant and Uddin (2015) found that for the DAN, children exhibited greater within network connectivity (short-range functional connectivity) in comparison to adults. In adults, long-range functional connectivity between DAN and regions outside the network is believed to enable greater top-down attentional capacities in adulthood (Rubia, 2013). For VAN, children showed greater functional connectivity than adults (Farrant & Uddin, 2015). The authors speculated that this over-connectivity in the VAN can perhaps explain why children are susceptible to interruption by environmental stimuli and are less able to maintain activities requiring top-down attentional control (Gaspelin, Margett-Jordan & Ruthruff, 2015; Bunge, Dudukovic, Thomason, Vaidya & Gabrieli, 2002).

The frontoparietal regions in the DAN are believed to be sources of attentional biases onto the visual cortex (Desimone & Duncan, 1995; Corbetta & Shulman, 2002; Reynolds & Chelazzi, 2004). Therefore, since spatial suppression surrounding the focus of attention is a result of top-down modulation, we hypothesized that only young adults, adolescents and perhaps pre-adolescents would exhibit attentional surround suppression. To test our hypothesis, we examined whether the separation distance between a spatially attended target and second target affected visual discrimination across development. Cutzu and Tsotsos' (2003) psychophysical task was replicated in Experiment 1 with participants between the ages of 8 years to young adulthood (18+ years). Accuracy was expected to increase as a function of the separation distance between the targets for age groups exhibiting attentional surround suppression. In our control experiment, Experiment 2, an independent group of participants were tested with a central cue to assure that our findings in Experiment 1 were in fact related to spatial attention. In Experiment 3, we tested an independent group of children with slower task parameters to afford them with a more feasible task and to examine whether their top-down processes need more time to tune their visual system.

## Materials and Methods

### Participants

For Experiment 1, 2 and 3, participants were recruited and tested at the Ontario Science Centre. For our young adult groups, Undergraduate Research Participant Pool students were also recruited to participate in the study. The general demographic information of all age groups in all three experiments is presented in Table 1.

### Stimuli and Task

*Experiment 1*

This experiment assessed whether a ring of suppression surrounding an attended item is observed in young adults and younger age groups. We replicated the first experiment of Cutzu and Tsotsos' (2003) psychophysical study, but with both young adults and developmental age groups. Participants aged between 8 and 27 years ($n = 180$) were required to detect two red letter character targets (Target 1 and Target 2) from among black letter distractors and report whether the targets were identical (L-L and T-T) or different (L-T or T-L). Participants' spatial attention was cued to one of the two letter targets (Target 1). The spatial cue focusing attention to one of the targets was expected to not only enhance the processing of the cued target but also suppress surrounding stimuli. Visual discrimination was therefore expected to improve as a function of inter-target separation, that is, the distance between Target 1 and Target 2, as a consequence of a lessening of spatial surround suppression as the distance from the attended location increased.

 The experimental sequence began with the cue, a light gray disk, which was briefly displayed and anticipated the location of the first target. The cue was presented for a duration of 100 msec and was valid on all trials. Following the cue, the visual array was displayed and consisted of 6 randomly oriented Ls and 6 randomly oriented Ts, arranged in the shape of a circle centered on a fixation point at the center of the screen. The radius of the circle was 4º and the character size was 0.6º visual angle. The items in the visual array were displayed in a circle to make sure that all items have equivalent retinal resolution. The letter characters were equally spaced out and were overlaid on top of a circular light disk, identical in size and colour to the cue disk. Two of the letter characters were red, one of which was cued target, Target 1, whose location was cued, while the remainder of the characters were black. The distances between the two target letters, Target 1 and Target 2 varied among six values of inter-target separation distances. The inter-target separation distances varied from where targets were neighbours, to where two targets were diametrically opposite, with five distracter characters between them. The inter-target distances were measured as a line segment between Target 1 and Target 2. At the largest inter-target separation distance, the distance was considered as 1.00. The smaller inter-target distances were considered as a fraction of the largest inter-target distance that it represents. The orientation of the line segment connecting Target 1 and Target 2 was random across all trials. Figure 1A and 1B depict examples of the six inter-target separation distances included in this experiment and the temporal sequence of a trial, respectively.

 Participants were given 3 blocks of practice trials. For the first block, the visual array was on for 500 msec, for the second 250 msec and finally for the third 175 msec. The decreasing duration of the visual array presentation during practice was found in pilot testing to greatly help younger age groups understand the task. In order to maintain consistency among all age groups, older participants, including adults, also underwent the practice blocks and were instructed in a similar manner to the younger groups.

Participants completed a total of 144 trials, in which each six inter-target separations were presented a total of 24 times, with 12 of those times being in the identical targets condition (LL or TT, 6 times each) and 12 times in the different targets condition (LT or TL, 6 times each). Trials were divided into 4 blocks. This provided the participants a short break in between each block and assured that all the participants remain focused on the task throughout the entire experiment. During the pilot phase of this study, a group of 6- to 7-year-olds were tested (n = 18), but they were unable to properly complete the task (e.g., could not complete all blocks, could not maintain focus, etc.) and were therefore excluded from the final study.

*Experiment 2*

In this experiment, an independent group of participants aged between 8 and 23 years (n = 164) were tested on a similar paradigm as in Experiment 1, with the exception of a central cue being presented instead of a spatial cue. This experiment was included to verify whether the results of Experiment 1 were in fact a consequence of spatial attention. All age groups were tested in separate sub-experiments.

*Experiment 3*

In Experiment 3, yet another independent group of 8- to 11-year-olds (n = 57) were tested on a modified version of the Experiment 1 paradigm, where the cue presentation time was doubled. All other task parameters remained the same as the Experiment 1 task. Experiment 3 allowed us to examine whether top-down feedback processes in 8- to 11-year-olds require more time in order to optimize the visual processing of attended stimuli and suppress the processing of surrounding stimuli. Both age groups were tested in separate sub-experiments.

**Results**

**Experiment 1**

Visual discrimination accuracy increased as a function of inter-target separation only in the 12- to 22-year-olds but not in 8- to 11-year-olds, suggesting that spatial suppression surrounding the focus of attention is only observed in the older developmental age groups. However, unlike in young adults where accuracy gradually increased as a function of inter-target separation, accuracy in the younger participants aged between 12 and 17 years did not increase until the largest separations of 0.97 and 1.00. This finding is surprising given that it suggests that the suppressive surround may encompass a larger area in 12- to 17-year-olds. The 8- to 11-year-olds did not exhibit any differences in accuracy across inter-target separation. Figure 2A depicts each age group's mean visual discrimination accuracy across inter-target separation for Experiment 1.

*Young Adults (18-22 years)*

In young adults, accuracy improved with increasing inter-target separation, increasing from approximately 60% when the targets were immediately adjacent to about 72% when diametrically opposite. A repeated-measure analysis of variance (ANOVA) was conducted using the linear mixed-effects function in R statistical software (R Core Team, 2013). Inter-target separation was set as a fixed variable and subject as a random variable. The main effect of inter-target separation on accuracy was significant, $F(5,135) = 11.33$, p < .0001. Bonferroni corrected post-hoc tests revealed that adults' accuracy was significantly lower at the minimum inter-target separation of 0.26 (M = .60, SD = .07) compared to separations of 0.71 (M = .67, SD = .10), 0.87 (M = .70, SD = 0.11), 0.97 (M = .71, SD = .11) and 1.00 (M = .72 SD = .11) (*p* < .001 for 0.26 compared to 0.71 and $p < .0001$ for all other comparisons). Adults' accuracy was also lower at inter-target separation 0.50 (M = .62, SD = .11) compared to 0.87 (M = .70, SD = 0.11), 0.97 (M = .71, SD = .11) and 1.00 (M = .72 SD = .11) ($p < .01$).

To further examine the hypothesis that accuracy is affected, and in fact improves as a function of inter-target separation, a linear regression analysis of the dependence of accuracy on inter-target separation was performed. The linear regression model was significant $F(5,162) = 6.20$, $p < .0001$, indicating that the null hypothesis of all the slope coefficients being equal to 0 can be rejected. In young adults, accuracy therefore increased as a function of inter-target separation. The R-squared statistic of the linear regression model was $R^2 = 0.16$, which as an index of effect size represents a medium effect (Cohen, 1988).

*Older Adolescents (16-17 years)*

Accuracy in 16- to 17-year-olds improved with increasing inter-target separation, increasing from approximately 58% when the targets were immediately adjacent to 70% when diametrically opposite. The repeated-measures ANOVA revealed a significant main effect of inter-target separation on accuracy, $F(5,150) = 9.50$, $p < .0001$. Bonferroni corrected post-hoc tests revealed that the 16- to 17-year-olds' accuracy was significantly lower at the minimum inter-target separation of 0.26 (M = .58, SD = .12) compared to separations of 0.97 (M = .69, SD = .13) and 1.00 (M = .70 SD = .12) ($p < .0001$). Accuracy was lower at inter-target separation 0.50 (M = .59, SD = .11) compared to 0.97 (M = .69, SD = .13) and 1.00 (M = .70 SD = .12) ($p < .001$). Accuracy was also lower at 0.71 (M = .59, SD = 0.13) compared to 0.97 (M = .69, SD = .13) and 1.00 (M= .70 SD = .12) (all $p$-values $< .001$).

The linear regression model was significant $F(5,180) = 6.12$, $p < .0001$, indicating that the null hypothesis of all the slope coefficients being equal to 0 can be rejected. In 15- to 16-year-olds, accuracy therefore increased as a function of inter-target separation. The R-squared statistic of the linear regression model was $R^2 = 0.15$, which as an index of effect size represents a medium effect (Cohen, 1988).

*Younger Adolescents (14-15 years)*

Accuracy in 14- to 15-year-olds improved with increasing inter-target separation, increasing from 60% when the targets were immediate neighbours to about 69% when diametrically opposite. The repeated-measures ANOVA revealed a significant main effect of inter-target separation on accuracy, $F(5,120) = 9.32$, $p < .0001$. Bonferroni corrected post-hoc tests revealed that participants' accuracy was significantly lower at the minimum inter-target separation of 0.26 (M = .59, SD = .09) compared to separations of 0.97 (M = .72, SD = .10) and 1.00 (M = .69 SD = .09) (p < .001). Accuracy was lower at inter-target separation 0.50 (M = .57, SD = .10) compared to accuracy at 0.97 (M = .72, SD = .10) and 1.00 (M = .69 SD = .09) (p < .05). Accuracy was lower at 0.71 (M = .64, SD = .11) compared to 0.97 (M = .71, SD = .10). Accuracy was also lower at 0.87 (M = .61, SD = 0.13) compared to 0.97 (M = .72, SD = .10) and 1.00 (M = .69 SD = .09) (*p < .05*).

The linear regression model was significant $F(5,120) = 7.85$, $p < .0001$, indicating that the null hypothesis of all the slope coefficients being equal to 0 can be rejected. In 14- to 15-year-olds, accuracy therefore increases as a function of inter-target separation. The R-squared statistic of the linear regression model was $R^2 = 0.25$, which as an index of effect size represents a medium to large effect (Cohen, 1988).

*Pre-Adolescents (12-13 years)*

Accuracy in 12- to 13-year-olds improved with increasing inter-target separation, increasing from 54% when the targets were immediate neighbours to about 65% when diametrically opposite. The repeated-measures ANOVA revealed a significant main effect of inter-target separation on accuracy, $F(5,175) = 7.26$, $p < .0001$. Bonferroni corrected post-hoc tests revealed that the 12- to 13-year-olds' accuracy was significantly lower at the minimum inter-target separation of 0.26 (M = .54, SD = .10) compared to separations of 0.97 (M = .63, SD = .14) and 1.00 (M = .65 SD = .11) ($p < .001$ for both comparisons). Accuracy was lower at inter-target separation 0.50 (M = .56, SD = .10) compared to of 0.97 (M = .63, SD = .14) and 1.00 (M = .65 SD = .11) (both at $p < .001$). Accuracy was also lower at 0.71 (M = .57, SD = 0.11) compared to 1.00 (M = .65 SD = .11) ($p < .01$).

The linear regression model was significant $F(5,210) = 5.27$, $p < .001$, indicating that the null hypothesis of all the slope coefficients being equal to 0 can be rejected. In 12- to 13-year-olds, accuracy therefore increased as a function of inter-target separation. The R-squared statistic of the linear regression model was $R^2 = 0.11$, which as an index of effect size represents the lower bounds of a medium effect (Cohen, 1988).

*Older Children (10-11 years)*

Accuracy in 10- to 11-year-olds remained at around 55% (range = 52% to 59%) and did not improve with increasing inter-target separation. The repeated-measures ANOVA showed no significant main effect of inter-target separation on accuracy, $F(5,140) = 1.81$, $p > .05$.

The linear regression model was not significant $F(5,168) = 1.23$, $p > .05$, indicating that the null hypothesis of all the slope coefficients being equal to 0 could not be rejected. The R-squared statistic of the linear regression model was $R^2 = 0.04$.

*Younger Children (8-9 years)*

Accuracy in 8- to 9-year-olds remained at around 53% (range = 51% to 55%) and did not improve with increasing inter-target separation. The repeated-measures ANOVA showed no significant main effect of inter-target separation on accuracy, $F(5,150) = 0.58$, $p > .05$.

The linear regression model was not significant $F(5,150) = 1.80$, $p > .05$, indicating that the null hypothesis of all the slope coefficients being equal to 0 cannot be rejected. The R-squared statistic of the linear regression model was $R^2 = 0.01$.

**Experiment 2**

In Experiment 2, when the cue was presented centrally and no longer cued attention to one of the targets, accuracy was not affected by inter-target separation in the age groups of pre-adolescents through young adults. This strongly suggests that the spatial suppression exhibited by participants in Experiment 1 was related to the focus of attention. Figure 2B depicts each age group's mean visual discrimination accuracy across inter-target separation in Experiment 2.

*Young Adults (18-23 years)*

Similar to Experiment 1, a repeated-measure analysis of variance (ANOVA) was conducted using the linear mixed-effects function in R. Accuracy did not change across inter-target separation. The main effect of inter-target separation on accuracy was not significant, $F(5,90) = 2.01$, $p > .05$.

A linear regression analysis of the dependence of accuracy on inter-target separation was also performed to examine whether there was a linear relationship between accuracy and inter-target separation. The linear regression model was not significant $F(5,108) = 1.07$, $p > .05$, indicating that the null hypothesis of all the slope coefficients being equal to 0 cannot be rejected

*Older Adolescents (16-17 years)*

Accuracy did not change across inter-target separation in the 16- to 17-year-olds. The repeated-measures ANOVA revealed no significant main effect of inter-target separation on accuracy, $F(5,115) = 1.93$, $p > .05$.

The linear regression model was not significant $F(5,138) = 1.51$, $p > .05$, indicating that the null hypothesis of all the slope coefficients being equal to 0 cannot be rejected.

*Younger Adolescents (14-15 years)*

The repeated-measures ANOVA revealed a significant main effect of inter-target separation on accuracy, $F(5,135) = 2.84$, $p < .05$. Bonferonni corrected post-hoc tests revealed that the main effect was only driven by the significantly lower accuracy at 0.26 (M= .51, SD = .07) in comparison to 1.00 (M = .59, SD = .11). No other inter-target separation accuracy comparison was significant.

Importantly, the linear regression model was not significant $F(5,162) = 2.15$, $p > .05$, indicating that the null hypothesis of all the slope coefficients being equal to 0 cannot be rejected. Accuracy in the 14- to 15-year-olds, therefore, did not increase as a function of inter-target separation.

*Pre-Adolescents (12-13 years)*

Accuracy did not increase as a function of inter-target separation in the 12- to 13-year- olds. The repeated-measures ANOVA revealed no significant main effect of inter-target separation on accuracy, $F(5,150) = 1.61$, $p > .05$.

The linear regression model was not significant $F(5,180) = 1.39$, $p > .05$, indicating that the null hypothesis of all the slope coefficients being equal to 0 cannot be rejected.

*Older Children (10-11 years)*

Accuracy did not increase as a function of inter-target separation in the 10- to 11-year-olds. The repeated-measures ANOVA revealed no significant main effect of inter-target separation on accuracy, $F(5,180) = 0.54$, $p > .05$.

The linear regression model was not significant, $F(5,216) = 0.51$, $p > .05$, indicating that the null hypothesis of all the slope coefficients being equal to 0 cannot be rejected.

*Younger Children (8-9 years)*

Accuracy did not increase as a function of inter-target separation, and performance was at floor in the 8- to 9-year-olds. The repeated-measures ANOVA revealed no significant main effect of inter-target separation, $F(5,120) = 1.10$, $p > .05$.

The linear regression model was not significant $F(5,144) = 0.99$, $p > .05$, indicating that the null hypothesis of all the slope coefficients being equal to 0 cannot be rejected.

## Experiment 3

Doubling the cue presentation time did not lead to spatial suppression surrounding the focus of attention in 8- to 11-year-olds, suggesting that the lack of surround suppression at these ages in Experiment 1 was not due to insufficient time for attentional feedback processes to have an impact. Figure 2C depicts each age group's mean visual discrimination accuracy across inter-target separation.

*Older Children (10-11 years)*

Accuracy did not increase as a function of inter-target separation. A repeated-measure analysis of variance (ANOVA) was conducted using the linear mixed-effects function in R. The main effect of inter-target separation on accuracy was not significant, $F(5,120) = 1.10$, $p > .05$.

A linear regression analysis of the dependence of accuracy on inter-target separation was also performed to examine whether there was a linear relationship between accuracy and inter-target separation. The linear regression model was not significant $F(5,144) = 0.99$, $p > .05$, indicating that the null hypothesis of all the slope coefficients being equal to 0 cannot be rejected.

*Younger Children (8-9 years)*

Accuracy did not increase as a function of inter-target separation in the 8- to 9-year-olds and they again performed close to floor (at chance – 50%). The main effect of inter-target separation on accuracy was not significant, $F(5,130) = 1.20$, $p > .05$. The linear regression model was not significant $F(5,156) = 1.13$, $p > .05$, indicating that the null hypothesis of all the slope coefficients being equal to 0 cannot be rejected.

**Discussion**

In adulthood, it is well established that attentional feedback processes impact visual processing by modulating activity in the visual cortex (Hopf et al., 2012). Visual cortext activity modulation occurs due to top-down attentional selection pruning and suppressing forward-projecting units or neurons not representing relevant input, which as a consequence gives rise to suppression surrounding the focus of attention (Tsotsos, 2005). In development, attention is even more critical because it is a time period during which an immense amount of learning and psychological change is taking place. Understanding the development of attention and more specifically the development of top-down attentional projections is therefore important to better understand how the typically developing brain processes visual information. The current study examined whether attentional surround suppression, a predicted by-product of top-down attentional modulation, is observed across a wide developmental age range. The current findings show that spatial attention similarly influences visual processing in late development. Spatial suppression surrounding the focus of attention was observed in young adults, adolescents and pre-adolescents, as predicted by studies of top-down attentional development.

According to the ST model (Tsotsos, 1995), selective attention is viewed as a process of winner-take-all (WTA), whereby a global winner is computed across the entire visual field and all of the connections of the visual pyramid that do not contribute to the winner are pruned. As a result, the selected stimulus in the input layer, for instance the spatial location of the cued target for Experiment 1, re-propagates through the network and is processed by the neurons without surrounding distracting stimuli. The eliminated or pruned projections of the neurons not representing the selected target stimulus form the suppressive surround. In the current study, not only did ST allow for an examination of top-down attentional development, it could also be use to correctly predict that pre-adolescents to young adults, whose top-down attentional mechanisms are nearly mature or mature, would exhibit suppression surrounding the focus of attention.

The lack of an inter-target separation effect on accuracy when a central target was used (Experiment 2), confirmed that our findings of surround suppression when a spatial cue was used (Experiment 1), were indeed related to spatial attention. In Experiment 2, a centrally presented cue lead to the suppressive surround manifesting around the center of the screen. Therefore, the targets and distractors would be equally partially suppressed, and suppression would thus not vary across inter-target separation. In Experiment 1, when the spatial cue focuses attention to one of the targets, enhanced processing of the cued target is accompanied by a suppressive surround. Therefore, when the second target is presented close to the attended target, as in case of inter-target separation 0.26 and the targets are side by side, it falls in the suppressive surround and becomes difficult to visually discriminate.

**Neural Development and Visual Attention**

As previously discussed, in early development, visual feedforward and low-level orienting mechanisms are thought to be more dominant, while top-down feedback processes continue to be strengthened (Amso & Scerif, 2015). That in the current study attentional surround suppression was only observed in the young adults and older developmental age groups is therefore not particularly surprising. In adults, long-range functional connectivity between the dorsal attention network (DAN), a neural network activated when top-down attention is focused, and regions outside the network is believed to enable greater top-down attentional capacities (Rubia, 2013). The lack of surround suppression in the 8- to 11-year-olds is therefore likely a consequence of immature top-down feedback projections that are not as strongly connected to further afield cortical regions at these ages. Indeed previous research has demonstrated that in children under the age of 12 years, the DAN is not as functionally connected to farther regions such as the visual cortex (Farrant & Uddin, 2015).

Studies examining the maturation of structural connectivity, that is the physical connections of long-range connections formed by white matter tracts (Khundrakpam, Lewis, Zhao, Chouinard-Decorte & Evans, 2016), have also shown that the maturity of structural connectivity is protracted, continuing into adulthood. In a longitudinal study, Lebel and Beaulieu (2011) used diffusion tensor imaging (DTI) to examine developmental changes in white matter in healthy participants aged from 5 to 32 years. Continued maturation was observed from childhood to adulthood for all 10 major white matter tracts, but notably, maturation of the inferior and superior longitudinal and frontal-occipital fasciculi continued into the twenties (Lebel & Beaulieu, 2011). These association tracts connecting the frontal areas to other brain regions support complex cognitive function such as inhibition, executive function and importantly, attention (Lebel & Beaulieu, 2011; Moll, Zahn, de Oliveira-Souza, Krueger & Grafman, 2005; Blakemore and Choudhury, 2006; Jung and Haier, 2007). In the context of the current study, it can therefore be speculated that these DTI findings support the idea that developmental differences in the manifestation of attention-modulated surround suppression are related to reduced connectivity between frontal brain areas and other regions of the brain.

The changes in white matter and connectivity from childhood to adulthood are believed to reflect increases in myelination and the axonal density (Khundrakpam et al., 2016). Cortical myelination occurs initially in the sensory tracts, followed by the motor tracts and finally the association tracts (Huttenlocher, 2002). White matter volume continues to increase with age during childhood and adolescence, and even continuing through adulthood (Lebel & Beaulieu, 2011), and importantly, the rate of volume increase varies by brain regions. For instance, in development, white matter increases in the occipital cortex are about 2.14% per year, whereas increases in the frontal cortex are only about 1.37% per year (Sowell, Peterson, Thompson, Welcome, Henkenius & Toga, 2003). This suggests that while white matter integrity in the sensory regions may be adult-like earlier in development, it takes far longer for white matter to completely mature in the frontal cortex, which in turn would likely affect the efficiency of top-down feedback modulation in development.

Indeed, white matter volume and myelination gain, particularly within frontal regions, has been found to be associated with improvements in cognitive processes (Khundrakpam et al., 2016). For instance, white matter volume in the frontal-striatal circuits is associated with better inhibitory control (Liston, Watts, Tottenham, Davidson, Niogi, Ulug & Casey, 2006). The fronto-striatal circuit is also believed to play a significant role in mediating attention (Wu, Gau, Lo & Tseng, 2012). Myelination facilitates interactions between brain regions, which leads to more efficient recruitment of the target neural population (Knyazeva, Fornari, Meuli & Maeder, 2006). Reduced myelination in the younger age groups, particularly in the frontal regions, therefore, can possibly lead to less efficient signal propagation from the frontal areas to the visual areas, resulting in less attentional modulation. Reduced attentional modulation would lead to reduced or no attentional surround suppression, which is what was indeed observed in the 10- to 11-year-olds and 8- to 9-year-olds of the current study.

But, for the pre-adolescents and adolescents, why did they exhibit a greater area of spatial suppression surrounding their focus of attention in comparison to adults? In adolescence functional activation is more spatially diffuse across frontal and parietal regions, whereas in adults activation is more focal and fine-tuned within the fronto-parietal network (Konrad, Neufang, Thiel, Specht, Hanisch, Fan, Herpertz-Dahlmann & Fink, 2005; Durston, Davidson, Tottenham, Galvan, Spicer, Fossella & Case, 2006). In adulthood, focal instead of diffuse activation is believed to represent reorganization in cortical areas, allowing for more efficient processing (Ungerleider, Doyon & Karni, 2002). In development, a change towards more focal functional activation is believed to be a result of synaptic pruning, which improves the signal to noise ratio in the neural system and strengthens relevant connections (Durston, Davidson, Tottenham, Galvan, Spicer, Fossella & Casey, 2006). Perhaps in the current study, a greater area of attentional surround suppression was observed in pre-adolescents and adolescents because functional connectivity between their frontal regions and visual cortex is not focal but rather more diffuse. Unlike in adulthood, attentional modulation of visual cortex activity in adolescence would therefore not be as specific and focal, and as a consequence, surround suppression would unnecessarily span over a larger spatial region.

**Development of Top-Down Attention**

Our findings converge well with previous research revealing a protracted maturation of top-down attention mechanisms. Visual search studies, for instance, have shown that despite bottom-up attentional mechanisms maturing early in development (Adler & Orprecio, 2005; Donnelly et al., 2007; Merrill & Conners, 2013; Taylor et al., 2003; Trick & Enns, 1998; Woods et al., 2013), top-down mechanisms are still developing in childhood (Donelly et al., 2007; Trick & Enns, 1998; Woods et al., 2013). The maturation of executive attention, the process of resolving conflict between competing inputs for the purpose of a goal driven task (Posner & Petersen, 1990), is also slow. Executive attention does not become more adult-like until around 14 years of age (Luna, Garver, Urban, Lazar & Sweeney, 2000). Our findings provide further support to the interpretation of late developing top-down attentional processes, by showing that, surround suppression, a predicted by-product of top-down attentional modulation on visual processing, is not present in children under the age of 12 years.

Models of visual attention development have proposed that early in development visual feedforward and low-level orienting mechanisms are more dominant, while top-down feedback processes are strengthened throughout development (Amso & Scerif, 2015, Atkinson, 2000; Johnson, 1990). Consequently, in younger age groups, feedfoward mechanisms are believed to be more heavily relied upon (Amso & Scerif, 2015), which can account for why the children in the current study did not exhibit attention surround suppression, even when their attention mechanisms were given more time to tune their visual system in Experiment 3. Importantly, an over-reliance on feedfoward processes, can also explain other development findings. For instance, children tend to be more susceptible to interference and less able to inhibit responses in comparison to young adults (Bunge et al., 2002). As previously discussed, the VAN, an attention network activated in cases where bottom-up processing is taking place, shows greater functional connectivity in children in comparison to adults (Farrant & Uddin, 2015). Over-activity in the VAN could also account for the increase in distractibility and disruption by environmental stimuli observed in childhood (Bunge et al., 2002). One possibility as to why a reliance of feedforward processes in children is beneficial or necessary at younger ages is that it allows for the detection of salient stimuli, important for survival (Farrant & Uddin, 2015). Throughout development, as top-down feedback processes mature, greater top-down attentional modulation takes place.

**Conclusions**

Having a better understanding of when and how attentional mechanisms develop and its effects on visual processing in development, is not just of theoretical importance, it also has practical relevance. For instance, from an educational perspective, highly decorated classrooms have been found to negatively impact children's learning, presumably because they are unable to inhibit salient distractors (Fisher, Godwin & Seltman, 2014). Having a better understanding of when top-down attentional processes develop and how immature attentional mechanisms impact visual and cognitive processes can therefore have major pedagogical implications.

From a clinical perspective, pervasive neurodevelopmental disorders such as Autism Spectrum Disorder (ASD) have been found to not only cause social-communicative and behavioural impairments (DMS-5 - American Psychiatric Association, 2013), but also sensory anomalies (Ronconi et al., 2018). For instance, individuals with ASD have been reported to exhibit visual sensory overload (Grandin, 2009) and more interference from irrelevant distractors (Adams & Jarrold, 2012; Remington, Swettenham, Campbell & Coleman, 2009). In a recent study, Ronconi and colleagues (2018) examined whether visual sensory anomalies in ASD are partially due to differences in attentional surround suppression. Remarkably, similar to our current study findings, their psychophysical results showed that typically developing adolescents (mean age of 14) exhibit suppression surrounding their focus of attention. In comparison to the typically developing adolescents, the ASD adolescents exhibited weaker attentional surround suppression. In a second experiment, Ronconi et al. (2018) used dense-array electroencephalography (EEG) to examine the neurophysiological underpinnings of surround suppression in typically developing and ASD children (mean age of 11 and 12 years respectively). In the typically developing children, the N2, a part of the family of components that reflect attentional selection of relevant stimuli in space (Bocquillon, Bourriez, Palmero-Soler, Molaee-Ardekani, & Derambure & Dujardi, 2009) and time (Ronconi, Pincham, Cristoforetti, Facoetti & Szűcs 2016), was suppressed for targets appearing in the surround of the attentional focus. This attentional surround-modulated N2 effect was observed 300 msec after the attention probe. In contrast, the ASD children did not exhibit the N2 effect, highlighting their deficits in inhibiting visual information outside the focus of attention.

The 10- to 11-year-olds in the current study did not exhibit suppression surrounding their focus of attention. In Ronconi and colleagues' (2018) study, however, the typically developing children aged at around 11 years did exhibit suppressed N2 for targets appearing in the surround of their attentional focus. This finding would suggest that attention-modulated surround suppression is present in 11-year-olds, despite it not strongly being observed in our current study. However, due to reasonable practical reasons, the children in Ronconi et al.'s (2018) second electrophysiology experiment did not complete all the conditions featured in their first psychophysical experiment conducted with adolescents. Therefore, it is currently unclear whether in contrast to the current study findings, their 11-year-olds participants demonstrate attentional surround suppression psychophysically, as would be expected in older age groups.

Notably, another factor to consider is that in Ronconi and colleagues' (2018) study, the attentional surround-modulated N2 effect in the 11-year-olds was observed 300 msec after an attention probe. This raises the question of whether the temporal parameters used in our study made the tasks too difficult for the younger children to complete, admittedly a potential limitation of our current study. Increasing the cue time in Experiment 3 was meant to overcome this limitation by providing the younger participants with more time to complete their feedback processes, but instead, perhaps increasing the visual array duration is what is necessary to make the task more feasible. For instance, keeping the spatial cue duration at 100 msec and increasing the duration of the visual array from 175 msec to 250 msec would have perhaps been more appropriate for the younger children. This change could have arguably still provided the younger age groups with more time to complete their feedback processes. If the top-down feedback processes were elicited soon after the onset of the spatial cue, increasing the visual array time to 250 msec would allocate close to 300 msec for the top-down processes to complete by the response mask. After all, the attentional surround-modulated N2 effect in the 11-year-olds of Ronconi et al. (2018) study was observed 300 msec after the attention probe. In a subsequent study, therefore, increasing the visual array duration of the current task in younger age groups while monitoring eye movements to assure that they remain fixated at the center of the screen can have great empirical and theoretical value. This manipulation would allow for an examination of whether attention-modulated surround suppression can indeed be observed in younger age groups.

Other considerations include examining whether surround suppression would be observed in children with different stimuli properties, such as varying the size or salience of the visual array or the individual stimuli. There are no differences in receptive field size, eccentricity and visual field coverage in early and intermediate visual areas in children (5 to 12 years) and adults (Gomez et al., 2018). And, in the current study, the visual array fit in the parafovea, a region with no visual field coverage difference between adults and children. However, it is still possible that larger and more salient stimuli could have made the task more feasible for the younger children. Especially, since children up to 11 years of age show greater crowding effects, that is, impaired target recognition caused by surrounding contours, in comparison to adults (Jeon, Hamid, Maurer & Lewis, 2010).

Another possible future direction is to confirm the current study findings with other psychophysical tasks. This is important not only for validation purposes but also because a more appropriate task for younger age groups may reveal different findings. For example, the orientation discrimination task used in the magnetoencephalography (MEG) study by Hopf and colleagues' (2006) may be slightly simpler, since there is only 1 target. In their study, participants were required to search for a red target C among blue distractor Cs (presented in a quarter circle) and report its orientation. On half of the trials an attention probe was flashed at the center C. By comparing accuracy across the five target-to-probe distances, ranging from PD0 (target presented at the probed location) through PD4 (target presented 4 items away from the probe), attentional surround suppression could be examined. Pertinently, a near identical task was used in Ronconi and colleagues' (2018) study, where children and adolescents were tested, suggesting that it may indeed be a more developmentally appropriate task. Using this task with younger age groups could also allow for further examination of the attentional profile of attention across development. Further, by using neuro-techniques, possible neurophysiological mechanisms underlying the developmental differences in attentional surround suppression could be uncovered. It would also be compelling to examine whether MEG results in adolescents, for example, would mimic the current psychophysical findings of greater suppression in this age group.

Overall, the current study results show that top-down attentional modulation affects visual processing in pre-adolescents and adolescents. With regard to attentional development and more specifically the development of top-down attention mechanisms, our findings provide further support for the notion that early in development visual feedforward and low-level orienting mechanisms are more dominant and that top-down feedback processes strengthen over the course of development (Amso & Scerif, 2015).

Attention is undoubtedly important because without our brain's ability to organize and filter relevant information from the overabundance of all available information, we would not be able to interpret and make sense of our environment. Attention is a gateway for information to access conscious perception and explicit memory (Shim, Alvarez & Jiang, 2008). In development, attention is likely even more critical because it is a period of time during which an immense amount of learning and psychological change is taking place. Understanding the development of attention and more specifically the development of top-down attentional projections is therefore important to the pursuit of understanding how the typically developing brain processes visual information. The current study is an important step demonstrating that top-down projections similarly affects visual processing in pre-adolescence, adolescence and young adults, while additionally highlighting how visual attention processes function differently in childhood.

**Acknowledgements**

We would like to thank Dr. Rachel Ward-Maxwell for her help in facilitating the set-up of our data collection at the Ontario Science Centre, and all the research assistants who assisted in the data collection for this study. This research was partially funded by the Hallward Fund of the Toronto Foundation.

**Table and Figures**

**Table 1.** General demographics information of participants in Experiment 1, 2 and 3.

| | Sub-Experiment Age Groups | Participants | Mean Age | Gender |
|---|---|---|---|---|
| **EXPERIMENT 1** (n = 180) | Young Adults | 28 | 19.75 (18.00-27.34) | Female = 17, Male = 11 |
| | Older Adolescents | 31 | 16.95 (16.05-17.84) | Female = 21, Male = 10 |
| | Younger Adolescents | 25 | 14.75 (14.10-15.89) | Female = 7, Male = 18 |
| | Pre-Adolescents | 36 | 12.80 (12.05-13.89) | Female = 15, Male = 21 |
| | Older Children | 29 | 10.76 (10.03-11.88) | Female = 6, Male = 23 |
| | Younger Children | 31 | 8.81 (8.01-9.90) | Female = 16, Male = 15 |
| **EXPERIMENT 2** (n = 164) | Young Adults | 19 | 23.31 (18.01-23.31) | Female = 10, Male = 9 |
| | Older Adolescents | 24 | 16.95 (16.17-17.96) | Female = 11, Male = 13 |
| | Younger Adolescents | 28 | 14.68 (14.05-15.93) | Female = 16, Male = 12 |
| | Pre-Adolescents | 31 | 12.85 (12.01-13.95) | Female = 16, Male = 11 |
| | Older Children | 37 | 10.61 (10.11-11.87) | Female = 16, Male = 21 |
| | Younger Children | 25 | 8.73 (8.12-9.99) | Female = 12, Male = 13 |
| **EXPERIMENT 3** (n = 57) | Older Children | 30 | 11.12 (10.08-11.97) | Female = 15, Male = 16 |
| | Younger Children | 27 | 8.72 (8.03-9.89) | Female = 18, Male = 9 |

Note. Participants = Number of participants included, Mean Age = Average age in years.

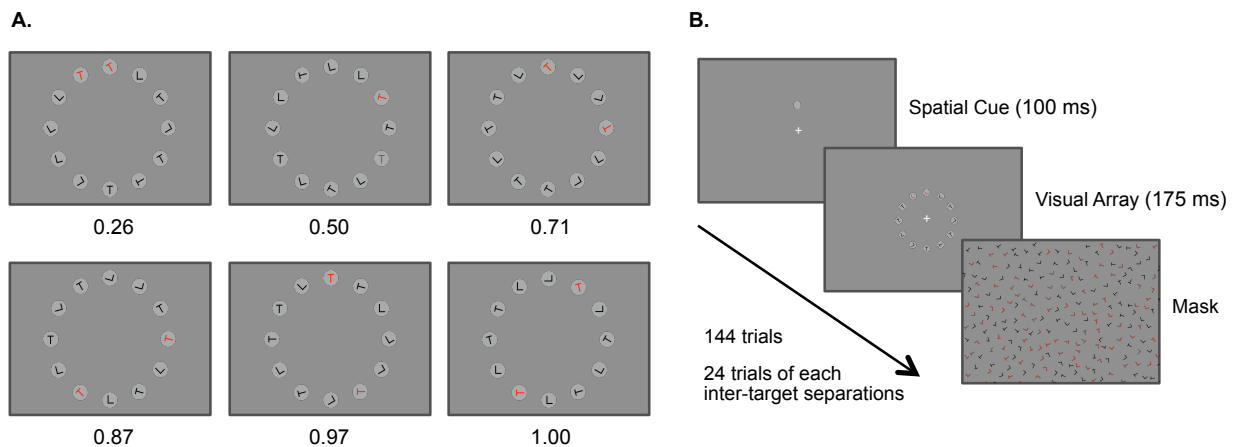

**Figure 1. A.** Inter-target separations included in the experiment. At the largest inter-target

separation distance, the distance was considered as 1.00. The smaller inter-target distances were considered as a fraction of the largest inter-target distance that it represents. **B.** Temporal sequence of Experiment 1.

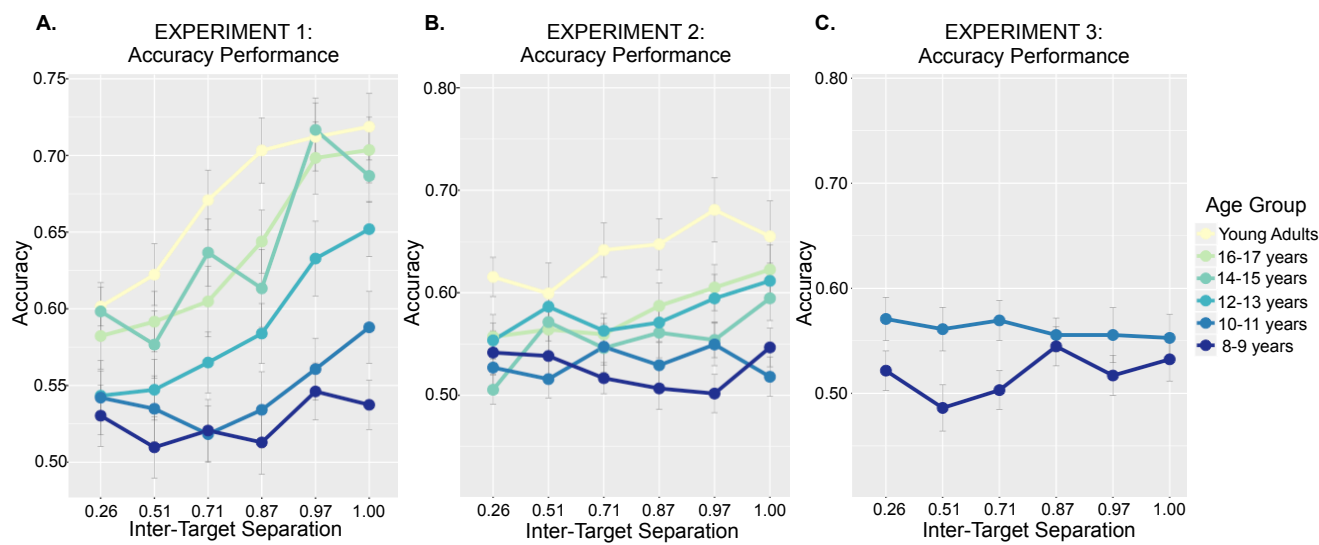

**Figure 2. A.** Visual Discrimination Accuracy of All Ages in Experiment 1. Visual discrimination

accuracies for each inter-target separations are depicted by age group. Visual discrimination

accuracy significantly increased as a function of inter-target separation in the 12 to 17 year-olds

and the young adults. However, in the 12- to 17-yearolds accuracy improvements were mainly

observed when the targets are largely separated such as for the inter-target separations of 0.97

and 1.00. Inter-target separation did not affect accuracy in the 8- to 11-year-olds. The error bars

indicate standard errors. **B.** Visual Discrimination Accuracy of All Ages in Experiment 2. Visual

discrimination accuracies for each inter-target separations are depicted by age group. Unlike in

Experiment 1, visual discrimination accuracy did not increase as a function of inter-target

separation. The error bars indicate standard errors. **C.** Visual Discrimination Accuracy of the 8-

to 11-year-olds in Experiment 3. Visual discrimination accuracies for each inter-target separations are depicted by age group. Visual discrimination accuracy was not affected by inter-target separation. The error bars indicate standard errors.